\documentclass[sigplan]{acmart}
\usepackage{tabularx}
\AtBeginDocument{%
  }

\setcopyright{acmlicensed}
\copyrightyear{2025}
\acmYear{2025}
\acmDOI{10.1145/XXXXXXX.XXXXXXX}

\acmConference[ICSE 2026 FoSE Track]{Future of Software Engineering}{May 2026}{Rio, Brazil}
\acmISBN{978-x-xxxx-xxxx-x/26/05}

\begin{document}

\title[Reclaiming Software Engineering]{Reclaiming
 Software Engineering as the Enabling Technology for the Digital Age}

\author{T. E. J. Vos, T. van der Storm, A. Serebrenik, L. Briand, R. Di Cosmo, J.-M. Bruel, B. Combemale}

\affiliation{
	\institution{Informatics Europe WG on Software Research}
	\country{}
}
\email{www.informatics-europe.org/research/software-research.html}

\renewcommand{\shortauthors}{Author et al.}

\begin{abstract}
Software engineering is the invisible infrastructure of the digital age. 
Every breakthrough in artificial intelligence, quantum computing, photonics, and cybersecurity relies on advances in software engineering, yet the field is too often treated as a supportive digital component rather than as a strategic, enabling discipline. 
In policy frameworks, including major European programmes, software appears primarily as a building block within other technologies, while the scientific discipline of software engineering remains largely absent.
This position paper argues that the long-term sustainability, dependability, and sovereignty of digital technologies depend on investment in software engineering research. 
It is a call to reclaim the identity of software engineering.
\end{abstract}



\maketitle

\section{Introduction}
The software engineering (SE) research community has never been larger or more productive. Conferences grow, journal and conference submissions multiply, and new research areas emerge at the intersection of SE with AI, quantum, photonics, and virtual worlds. Yet beneath the surface lies unease: our field’s identity is dissolving into the technologies it enables. We are indispensable, but invisible.

Historically, SE was understood as the science of building dependable (safe, secure, reliable), maintainable, and adaptive systems. Today, it risks being perceived as a commodity, tasked with supporting AI models, quantum devices, or virtual worlds, without being recognised as part of the scientific foundation that makes them possible.

This loss of identity is not only a visibility problem; it is a community problem. When SE is treated as a supporting activity rather than a scientific discipline, it becomes harder to define our shared challenges and attract and retain talent and funding. This position paper is a call to reclaim the identity of software engineering. This is a fundamental element of the discussion required to define the desired shape and role of the software engineering community going forward.

{\setlength{\parindent}{0pt}
\setlength{\parskip}{4pt plus 2pt minus 1pt}

\section{The erosion of identity}

We identify three forces that marginalise our field:

\noindent \textbf{Policy blindness.} In many funding programmes, software is treated as an auxiliary technology, or a supporting building block, rather than a key enabler. This misperception weakens long-term investments and pressures research toward short-term engineering rather than foundational inquiry.

\noindent \textbf{Fragmented incentives.} Internal metrics, the publish-or-perish culture, the quantification of research outputs, and stringent validation demands encourage incremental innovation. Expectations around validation have also become so rigid that they can overshadow the value of a well-motivated idea. 
Validation expectations can limit conceptual innovation, while quantifying research output can make it difficult to conduct industry-relevant and longitudinal studies.
Even when a concept is clearly articulated and its feasibility demonstrated, the pressure for exhaustive empirical proof discourages the early-stage and exploratory research the field urgently needs.

\noindent \textbf{Industrial disconnect.} Many of today’s most complex software systems are built inside organisations that do not primarily see themselves as software companies: semiconductor manufacturers, automotive, transport, healthcare, and financial institutions. One major Dutch bank put it plainly: ``We are essentially a software company that happens to understand money.'' Similar shifts are visible across sectors. Yet these sectors work largely in parallel, each developing its own tools, processes, and architectural conventions. Although we recognize significant diversity across application domains, without a shared, horizontal software engineering foundation, organisations repeatedly solve the same problems in isolation and reinvent practices that should be common. At the same time, researchers rarely gain access to the large, safety-critical, and embedded systems that define modern practice, while many industrial teams view academic work as too abstract or misaligned with their constraints. The result is a scientifically active field with limited impact.

\section{Software engineering enables everything}

Frontier technologies rely on software, not as a convenience but as the condition for scaling, integration, and safe operation. Across domains, progress slows when SE lags.

\begin{description}
\item[AI] Modern AI systems depend on reliable data pipelines, scalable and maintainable architectures, strong testing practices, monitoring infrastructures, and mechanisms for interpretability and accountability. This can only be achieved with mature software engineering solutions.

\item[Quantum and photonics] These fields require complex software stacks for simulation, compilation, control, calibration, and error mitigation. As devices grow in scale and precision, the software that orchestrates them becomes a dominant source of complexity. 
Hardware advances that are not matched by comparable progress in software engineering increasingly limit what these systems can deliver in practice.

\item[Cybersecurity] Security can only be achieved when systems are designed, implemented, and maintained using sound SE principles. Vulnerabilities rarely stem solely from cryptography; they emerge from flawed architectures, unsafe interfaces, poor configuration management, and inadequate development practices.

\item[Virtual worlds]
and emerging Web~4.0 ecosystems depend on software to provide persistent, interoperable, and human-centred infrastructures.

\item[Sustainability] Reducing the energy and resource footprint of digital systems demands energy-aware design, efficient architectures, optimised execution environments, and long-term lifecycle management. These are core software engineering concerns.
\end{description}

Across all these areas, SE provides the methods, abstractions, and infrastructures that make complex software systems dependable and beneficial in practice. Without continuous progress in software engineering, frontier technologies cannot mature, be deployed at scale, or meet societal expectations for safety, reliability, security, and sustainability.

\section{Reclaiming our identity in Europe}

Across Europe, national associations and research consortia have consistently raised concerns about the under-representation of software engineering in strategic research agendas. Recent position papers, open letters, and manifestos \cite{ercim, versen, ie, call_prioritise} all point to the same contradiction: the scientific identity of software engineering is fading at the very moment when well-engineered software is more critical than ever. 

In line with this, we have proposed a community-driven realignment of research, education, and policy to elevate software research as a strategic priority in Europe. A joint working group\footnote{https://www.informatics-europe.org/research/software-research.html}
between ERCIM and Informatics Europe has been established, with the long-term vision and agenda for:

\begin{itemize}
\item Recognising software engineering as essential digital infrastructure, fundamental to the scalability, reliability, and safety of all frontier technologies.

\item Establishing stable, long-term funding programmes for research that connects academia, industry, and the public sector.
Europe needs sustained investment in the scientific backbone of software: methods, tools, open-source ecosystems, and experimental research that is not tied to short-term application projects. This includes addressing industry challenges.

\item Ensuring that high-quality software is not only a technical requirement; it is a societal one. Responsible technology use, sustainability, reliability, long-term maintainability, and public trust must be embedded in the design and evolution of software systems. This requires research, training, and policy that reinforce these principles across sectors.
\end{itemize}

Reclaiming SE’s identity is not just important for our field. It is a strategic requirement for the future. Innovation in frontier technologies is only possible if we continue to invest in the SE foundations on which all these domains depend.

\section{Planning for change}

The ICSE Future of Software Engineering (FoSE) track explicitly asks how the community can shape the field's future. We propose five coordinated areas of action: using ICSE itself as a mechanism for reflection, strengthening the discipline's scientific core, aligning academia, industry, and policy around a shared research agenda, and making available to the field the long-overdue large-scale research infrastructure that is needed to address the challenges of today and tomorrow.

\subsection*{1. Use FoSE as a structured moment of collective reset}

The FoSE programme already invites the community to reflect on its values, publication culture, industry relationships, and reward structures. 
We think these topics only make sense when seen in light of the broader question of how SE defines and sustains its identity. 
The pre-event survey, workshop, and plenary debrief provide the appropriate structure to address these interconnected issues together and to formulate a shared vision for the community’s future.

\subsection*{2. Clearly position the scientific core of SE}

Conferences, journals, and steering bodies should articulate the long-term research challenges that define SE as a discipline, rather than allowing them to be overshadowed by application-driven trends. A clearly defined shared identity will help attract talent, improve research coherence, and strengthen our voice in policy discussions.

\subsection*{3. Consistent message across academia and industry.} Fragmented messages weaken our field, while a shared voice strengthens it.
If we want SE to regain a clear identity, our message must be consistent across academia and industry. Companies are better positioned to amplify awareness of high-priority challenges, including those listed below: 

\begin{itemize}
    \item Growing dependence on third-party platforms and the loss of digital autonomy.
    \item Concerns over the security and reliability of the (open-source) software supply chain.
    \item Legacy software and the lack of transferable skills across generations of systems.
    \item Navigating regulatory compliance in highly regulated sectors.
    \item Dealing with the rising complexity in systems: scale, variability, versioning.
    \item Responsibly integrating generative AI into software development workflows.
    \item Difficulty attracting and retaining talent.
\end{itemize}

 We call on the ICSE community, when working with industry partners, to ensure that these concerns are articulated clearly and coherently, and expressed as problems to be addressed by software research.

\subsection*{4.  Alignment through a shared research agenda}

The ERCIM/IE Working Group and its partners are currently developing a Strategic Research and Innovation Agenda (SRIA). 
It is intended to provide a clear, shared view of long-term research challenges in software engineering and to guide funding and policy decisions. 
The ICSE community can reinforce this effort by contributing to the SRIA, using it to inform its own roadmaps and reports, and by endorsing it as a standard reference for the field.

\subsection*{5. Invest in shared research infrastructure}

To support cumulative, reproducible, and policy-relevant research, the community should treat shared, open, and long-lived software engineering infrastructures as first-class research instruments. Persistent artifact identifiers, universal archives, reproducible build frameworks, and longitudinal datasets are essential for studying software systems at the scale and time horizons at which they now operate.

\section{Conclusion}
Software Engineering research and innovation have built the digital world, but it now faces an existential crisis. To remain relevant, we must reclaim our identity as a scientific discipline that enables the digital age worldwide.

\begin{acks}
We acknowledge all people who have been involved in the meetings and WGs we have constituted:

\begin{table}[h]
\centering
\begin{tabularx}{\columnwidth}{|p{2.6cm}|p{0.6cm}|X|}
\hline
VERSEN\footnote{\url{https://versen.nl}} & NL & 
Tanja Vos\\
& & Tijs van der Storm\\
& & Alexander Serebrenik\\
& & Paris Avgeriou\\
& & Michel Chaudron\\
\hline
GDR-GPL\footnote{\url{https://gdr-gpl.cnrs.fr}} & FR & 
Jean-Michel Bruel\\
\hline
INRIA\footnote{\url{https://inria.fr}} & FR & 
Benoit Combemale\\
& & Roberto Di Cosmo\\
\hline
SISTEDES\footnote{\url{https://www.sistedes.es}} & ES & 
Ernest Teniente\\
& & Elena María Navarro\\
& & Ernesto Pimentel\\
\hline
GRIN\footnote{\url{https://www.grin-informatica.it/chi-siamo/}} & IT & 
Filippo Lanubile\\
& & Azzurra Ragone\\
\hline
GII\footnote{\url{https://www.gii.it/index.php/gii/presentazione}} & IT & 
Luciano Baresi\\
& & Elisabetta Di Nitto\\
\hline
FB-SWT\footnote{\url{https://fb-swt.gi.de/}} & DE & 
Regina Hebig\\
& & Kurt Schneider\\
& & Jürgen Mottok\\
\hline
LERO\footnote{\url{https://lero.ie/}} & IE & 
Lionel Briand\\
& & Anthony Ventresque\\
\hline
CHOOSE\footnote{\url{https://choose.swissinformatics.org/}} & CH & 
Timo Kehrer\\
\hline
MDENet\footnote{\url{https://mde-network.com/}} & UK & 
Steffen Zschaler\\
\hline
SAST\footnote{\url{https://sast.se/}} & SE & 
Emil Alégroth\\
\hline
Software Center\footnote{\url{https://www.software-center.se/}} & SE & 
Malin Rosqvist\\
& & Mikael Sjödin\\
\hline
\end{tabularx}
\end{table}

\end{acks}

\bibliographystyle{ACM-Reference-Format}

\end{document}